\documentclass{llncs}
\usepackage{amssymb}

\title{An Application Specific Informal Logic for Interest Prohibition
       Theory}

\author{J.A. Bergstra \and C.A. Middelburg}
\institute{Section Theory of Computation, Informatics Institute,
           University of Amsterdam \\
           Science Park~904, 1098~XH Amsterdam, the Netherlands \\
           \email{J.A.Bergstra@uva.nl, C.A.Middelburg@uva.nl}
          }

\begin{document}

\maketitle

\begin{abstract}
Interest prohibition theory concerns theoretical aspects of interest
prohibition.
We attempt to lay down some aspects of interest prohibition theory
wrapped in a larger framework of informal logic.
The reason for this is that interest prohibition theory has to deal with
a variety of arguments which is so wide that a limitation to so-called
correct arguments in advance is counterproductive.
We suggest that an application specific informal logic must be developed
for dealing with the principles of interest prohibition theory.
\par\addvspace{1.5ex} \small {\sl Keywords:}
informal logic, interest prohibition, Islamic finance.
\nolinebreak
\end{abstract}

\section{Introduction}
\label{sect-intro}

Interest Prohibition Theory (IPT) concerns theoretical aspects of
interest prohibition.
Apart from any intrinsic merits of prohibiting interest to which one may
or may not subscribe, the importance of IPT is undeniable because of the
impact that various implications of interest prohibition have had, and
still have, on the design of systems of Islamic finance.%
\footnote
{The current existence and the remarkable success of systems of Islamic
 Finance has been documented in many sources.
 We mention~\cite{DP99a,ElG01a}.
 For the roots of Islam (and of Islamic IPT), we refer to~\cite{Don10a}.
}
The heroic battle fought by many countries against their excessively
large and interest bearing government debts is an additional reason to
consider the phenomenon of interest with caution, and perhaps to look
for its replacement by alternative mechanisms.
Difficulties caused by the interest mechanism at an individual level are
put forward in~\cite{Lew99a}.

We will attempt to lay down some aspects of IPT wrapped in a larger
framework of informal logic.
Doing so is needed because, unlike many conventional theories that one
finds in science and in the humanities, IPT needs to deal with a variety
of arguments which is so wide that a limitation to so-called correct
arguments in advance is counterproductive.
Informal logic is claimed to be topical by nature (see \cite{Kre08a})
and for that reason we suggest that an Application Specific Informal
Logic (ASIL) must be developed for dealing with the principles of IPT.%
\footnote
{The phrase ``application specific informal logic'' has been designed
 after the well-know phrase ``application specific integrated circuit''.
 For informal logic, see~\cite{Gro07a,Joh06a,Wal01a}.
}
The ASIL in question will be baptized ASIL$_\mathrm{IPT}$.

The division between IPT and its underlying ASIL requires some
explanation.
If an ASIL is left implicit, some general rational form of scientific
thinking that underlies most contemporary scholarly work is assumed.
Then many arguments about interest prohibition must be handled with
utmost care, and a distant style of writing emerges: ``A writes P about
Q'', ``A considers argument P convincing in the context of Q'', ``A
declares P to be a universal truth'', and so on.
An ASIL may allow some moderate import of Islamic or other religious
thought and still be acceptable to an average reader.

\section{Principles of Interest Prohibition}

IPT starts out from the observation that certain transactions have so
often been condemned in the past,%
\footnote{Mainly, but not exclusively in a religious setting.}
and in so many different cultures, that sustained prohibition may be
justified primarily on the basis of the combined authority of these
condemnations.
We will now discuss a key example of such transactions.

\subsection{Critically Productive Transactions}

We say that a transaction between parties A and B is Lender-side
Critically Productive (LCP) if A acquires goods, services or valuables
G in compensation of lending valuables V to B for some limited duration.
It is assumed that these valuables serve no instrumental purpose for B
or any other agent other than serving as a means of exchange or as a
store of value.
We note that:
(i)~the generation of G is the productive aspect, more specifically B is
productive;
(ii)~V may be understood as an asset which represents
wealth;
(iii)~B or B's sources cannot use V as a tool of some kind;
(iv)~the apparent circumstance that A is not involved in any substantial
fashion in the generation of G constitutes the critical aspect of the
transaction.

LCP transactions have often been forbidden and are still considered
problematic by many people.
Not only A's receipt of G, but also B's transferal of G should not take
place.
Contracts, agreements, promises or claims based on the expectation of
such receipts and transferals are considered undesirable as well.

Less prominent in IPT are transactions that are Borrower-side Critically
Productive (BCP).
In a BCP transaction, the borrower of valuables produces assets V
without sharing these with the owner of the valuables in any pre-agreed
fashion.

We incorporate in ASIL$_\mathrm{IPT}$ the assertion that LCP
transactions are forbidden.
We do not include the assertion that BCP transactions are forbidden,
though IPT experts will hold that as well.

\subsection{A Survey of Possible Viewpoints}

We have tried to formulate the point of departure independent of the
concept of money, hoping that a more general form is obtained in that
way. Here is a number of positions that one may hold towards the LCP
status of various kinds of transactions:
\begin{enumerate}
\item
LCP status assignment is community dependent. A most significant
community to consider is the Islamic Umma.
\item
All transactions involving interest are LCP transactions.
\item
Attributing LCP status to a (type of) transaction can be done by means
of an authoritative statement issued by a number of scholars from a
community.
\item
If an (expected) portion of income during a transaction makes it LCP,
the transaction can be saved from that status by the earner if he
promises to hand the income over to those in need of support immediately
after it is obtained.
\item
All interest based consumer credit transactions are LCP transactions.
\item
Savings account related transactions involving interest are not LCP.%
\footnote
{In~\cite{BM10b}, a detailed specification of a savings account is
 given.
 It appears that this matter is unexpectedly complex and that many
 options are left when it must be decided what exactly is to be
 forbidden once the savings account is labeled as LCP.
}
\item
LCP transactions are only those where an unlimited multiplication of
debt may occur, in particular the so-called doubling scheme: B owes A a
sum $s$ and at maturity date $t$, ending a period of length $p$, B can
either return $s$ to A or B can postpone repayment until $t + p$ but in
that case acknowledging a debt of size $2 \cdot s$.
\item
LCP transactions include at least the doubling scheme transactions.%
\footnote
{This follows from revealed sources.
 For an informative theory of authoritative revelations, we refer
 to~\cite{Whi10a}.
}
\item
All transactions involving excessive interest are LCP. Here interest is
considered excessive if its payment creates significant hardship on the
lender.
\item
Interests paid by a state on governmental bonds are not LCP.
\item
The boundary of LCP is flexible.
For that reason, the LCP prohibition is fundamentally open ended.
When new financial products are developed, a decision needs to be made
as to their non-LCP status.
Different bodies may disagree about that judgement.
\item
Micro-credits are not LCP because these are vital in conditions of
poverty.
\item
Many financial products known in the West, often called conventional
financial products, feature transfers of (concealed) interest at closer
inspection.
Such products may become classified LCP in due time.
\item
Sophisticated financial engineering often allows the development of a
non-LCP replacement of a product originally featuring concealed
interest.
\end{enumerate}

\section{Contemporary Interest Prohibition Theory}
\label{sect-contemporary-IPT}

Nowadays only Islam pronounces LCP status of financial products with the
intention of impeding their use.
For that reason, contemporary IPT is approximately the same as
Islam-based IPT.%
\footnote
{The development of logic for use within IPT has already been outlined
 in~\cite{Ber11a}.
}
We will now give a survey of the key assertions that contemporary IPT
holds.

\subsection{Assertions}

In this section, we distinguish, in addition to the LCP status, the
semi-LCP status: a product or transaction is semi-LCP if it is
problematic on LCP grounds, but the consequences of prohibition may be
too drastic so that a weaker judgement is asserted.

The assertions that contemporary IPT holds, and from which it can be
further developed, include the following:
\begin{enumerate}
\item
Financial products can be modelled as processes known from process
theory in computer science.%
\footnote
{For instance thread algebra \cite{BM04c} can be used for financial
 process description, see also \cite{BM10b}.
}
\item
The use of LCP products is prohibited.
\item
The use of semi-LCP products must be minimized both in number and in
size.
\item
The doubling debt scenario, which is nowadays outdated, is LCP.%
\footnote
{Has the doubling debt scenario become outdated because it was
 considered LCP already long ago?
}
\item \label{AbdReas}
The extension of the class of LCP products from the doubling scheme to a
comprehensive inclusion of interest related products has been put
forward by Abu Bakr al-Jassas in 981~AD (see~\cite{Far07a}).
This step involves abductive reasoning.%
\footnote
{One may conjecture that this extension of the LCP classification has
 been triggered by an increased popularity of ancient Greek philosophy,
 where interest prohibition is also found. Interest prohibition may have
 been accepted in ancient times because of a lack of experience.
 Unconstrained interest prohibition resulting from a LCP classification
 of all products involving interests was  found ``mainstream'' and for
 that reason unproblematic by al-Jassas.
 The exegetical strategy that revealed sources are likely to contain
 special cases for which adequate generalizations are to be found by
 means of scholarly work suggested al-Jassas to take this extension
 (with respect to the doubling scheme) on board, without feeling a need
 to consider possible disadvantages of that step.
 This explanation of how universal interest prohibition may have come
 about may be read as a form of conjectural history, a well-known style
 in writings on the history of money.
}
\item
In conventional financial systems, it is often the case that certain
financial products are considered inadmissible because of excessive
interest claims.
All such products are LCP.
\item
Some contemporary financial products are admissible in conventional
financial systems, but at the same time LCP.%
\footnote
{For that reason, Islamic Finance differs from conventional finance.}
\item
All products featuring interest are semi-LCP.
\item
The net social effect of preventing or minimizing the use of financial
products with LCP status is positive.
\item
For all LCP products, a non-LCP replacement product can be developed.
\item
Non-LCP replacement products are admissible in conventional financial
systems.
\end{enumerate}

\subsection{Open Issues}

The following questions can be considered open from the point of view of
IPT, though many adherents of Islamic Finance will claim that each reply
is positive:
\begin{enumerate}
\item
Are all forms of interest indications that a product is LCP?
\item
Is a positive interest bearing savings account LCP?
\item
Has a proof been established that such a savings account is LCP?
\item
If so, is the proof constructive from established sources?
\item
Is a positive interest bearing savings account LCP even if the rate
inflation exceeds the interest rate?
\item
Can conventional methods of econometrics and social sciences be applied
to demonstrate that avoiding LCP products has socially positive
consequences, at least from an Islamic perspective?
\item
If classification of a product as LCP leads to harm, because it cannot
be applied, is it nevertheless justified to maintain the LCP
classification?
\item
Can LCP classification be issued independently of a theory of
money?%
\footnote
{Important positions on IPT have been developed before modern fiat money
 came into existence.
}
\item
Is the following assertion valid: ``if a negative answer is given to one
or more of the above questions, that state of affairs will lead to useful
innovations of Islamic Finance''?%
\footnote
{As a material implication, which is the intended reading, this assertion
 is a valid one for those who give affirmative answers to the preceding
 questions.
 As a relevant implication, however, it is not.
}
\end{enumerate}
Different answers to these questions may give rise to different branches
of IPT.

\section{Tenets and Boundaries of ASIL$_\mathrm{IPT}$}

The methodological aspects for IPT that together constitute the core of
an ASIL for contemporary IPT include the following:
\begin{enumerate}
\item
Arguments for the LCP status of financial products or classes of
products can have different forms.
\item
LCP status judgements may change in time because consensus may change.
\item
LCP status requires proof, and in addition to that it may have an
economic rationale which indicates what injustice is prevented by
disallowing the application of a LCP labeled product.
\item
The use of analogies (see \cite{Mas98a})  plays no role in IPT.
\item
There is no application of inductive reasoning in IPT.
\item
There is one significant  application of abductive reasoning in IPT
(see item~\ref{AbdReas} in the enumeration of assertions in
Section~\ref{sect-contemporary-IPT}).
\item
Positions taken by scholars on matters of IPT are fixed once they pass
away. Subsequent determination of such positions  may be very difficult
due to lack of information.
\item
In some cases, authoritative judgement by later scholars may overcome
this lack of information -- for instance by making use of abduction.
\item
Determination of the position of scholars known to have been
contemplating IPT is a matter of ongoing work.
\item
It is conceivable in principle that scholars take or have taken
authoritative and definite positions in IPT which impact the
classification of future products that are entirely non-existent at the
time of their work.
\item
The rich design theory of non-LCP replacement products belongs to IPT.
\end{enumerate}
The possible forms of arguments for the LCP status of financial products
or classes of products include:
\begin{enumerate}
\item
Proof by conclusive argument from revealed sources in the absence of
opposing arguments in revealed sources.
\item
Proof by strong arguments from revealed sources in the presence of
contradictory arguments in revealed sources, priority of arguments
having been established authoritatively by those who have seen the
prophet alive, thereby producing derived sources.
\item
Reasoning by conclusive argument from derived sources, performed by
scholars having access to the full volume of derived sources, assigning
priority to particular arguments when needed, thereby producing
secondary derived sources.
\item
Consensus formed by authoritative opinions of a significant number of
informed scholars, who estimate the balance of a long tradition of
secondary derived sources by intuitive means.
\end{enumerate}

ASIL$_\mathrm{IPT}$ must also determine and explain the boundaries of
IPT.
Themes that are relevant in this respect include:
\begin{enumerate}
\item
The accommodation of the representation of dedicated completed versions
of IPT, including both versions that were found in the past and versions
that are operational at present.
\item
The presentation and explanation of the current versions of mainstream
IPT together with the major disputes that show there differences an that
drive their further development.
\item \label{theme-rationale}
The selection of overall rationales for interest prohibition.
\end{enumerate}
Theme~\ref{theme-rationale} requires an explanation.
In the conclusions of his master's thesis~\cite{Sub01a}, Azeemuddin
Subhani convincingly argues that such matters need thorough
investigation even if one takes the revealed sources as axiomatic
truths.
Doing so means little without a framework of interpretation and the
current framework leaves open too many questions.
One of the degrees of freedom in this matter is the proper
identification of the asset class of ``valuables'', on which the
definitions of LCP and BCP are grounded.
This asset class need perhaps not include modern bank money because that
money exists only in the form of debts which are necessarily either LCP
(if the bank pays interest to an account holder) or BCP (if no interest
is paid).
Perhaps access to assets from this class was considered so exclusive
that its possession imposed very special responsibilities on its owner.

Methodological questions can be posed as well.
These questions are matters that an ASIL may yet have to resolve before
being mature for the subject at hand:
\begin{enumerate}
\item
In what cases may or must universal claims that products with certain
properties are LCP be limited to a historically determined scope?
For instance, are there authors whose work cannot possibly be considered
to have been meant to be applied to something as remote as a credit
card?
\item
Disadvantages are claimed for non-LCP products replacing LCP products
(see~\cite[Section~4.3]{BM10b} and the website~\cite{ElD97a}).
Must replies to such claims be formulated?
\item
If a significant departure from established scholarly judgements must be
effected, what course of action might work?
Which modifications of contemporary IPT can still be imagined?%
\footnote
{As a thought experiment: what course of events might lead to a state of
 affairs where the doubling scheme is declassified as LCP? Or is that
 unimaginable?
}
\item
If in some time to come a significant departure from current IPT
viewpoints will be effected within what is then called Islam, must it
be the case that the degree of freedom that is made use of is already
detectible from the current state of knowledge?
\end{enumerate}

\bibliographystyle{splncs03}
\bibliography{IF}

\begin{thebibliography}{10}
\providecommand{\url}[1]{\texttt{#1}}
\providecommand{\urlprefix}{URL }

\bibitem{Ber11a}
Bergstra, J.A.: Real islamic logic. {\tt arXiv:1103.4515v1 [cs.LO]} (2011)

\bibitem{BM10b}
Bergstra, J.A., Middelburg, C.A.: Preliminaries to an investigation of reduced
  product set finance. To appear in Journal of King Abdulaziz University:
  Islamic Economics. Preliminary version: {\tt arXiv:1012.4291v1 [q-fin.GN]}
  (2010)

\bibitem{BM04c}
Bergstra, J.A., Middelburg, C.A.: Thread algebra for strategic interleaving.
  Formal Aspects of Computing  19(4),  445--474 (2007)

\bibitem{DP99a}
Dar, H.A., Presley, J.R.: Islamic finance: A western perspective. Journal of
  Islamic Financial Services  1(1),  3--11 (1999)

\bibitem{Don10a}
Donner, F.M.: Muhammad and the Believers: At the Origins of Islam. Harvard
  University Press, Belknap Press, Cambridge, MA (2010)

\bibitem{ElD97a}
{El Diwany}, T.: Islamic-finance.com. {\tt http://www.islamic-finance.com}
  (from 1997 onwards)

\bibitem{ElG01a}
El-Gamal, M.A.: An economic explication of the prohibition of \textit{Gharar}
  in classical {Islamic} jurisprudence. Islamic Economic Studies  8(2),  29--58
  (2001)

\bibitem{Far07a}
Farooq, M.O.: Stipulation of excess in understanding and misunderstanding
  \textit{riba}: The {al-Jassas} link. Arab Law Quarterly  21(4),  285--316
  (2007)

\bibitem{Gro07a}
Groarke, L.: Informal logic. Stanford Encyclopedia of Philosophy, {\tt
  http://plato.\linebreak[2]stanford.edu/entries/logic-informal} (2007)

\bibitem{Joh06a}
Johnson, R.H.: Making sense of ``informal logic''. Informal Logic  26(3),
  231--258 (2006)

\bibitem{Kre08a}
Kreuzbauer, G.: Topics in contemporary legal argumentation: Some remarks on the
  topical nature of legal argumentation in the continental law tradition.
  Informal Logic  28(1),  71--85 (2008)

\bibitem{Lew99a}
Lewison, M.: Conflicts of interest? {The} ethics of usury. Journal of Business
  Ethics  22(4),  327--339 (1999)

\bibitem{Mas98a}
Mas, R.: Qiy as: A study in {Islamic} logic. Folia Orientalia  34,  113--128
  (1998)

\bibitem{Sub01a}
Subhani, A.: The {Islamic} Doctrine of Riba Prohibition. Master's thesis,
  Institute of Islamic Studies, McGill University, Montreal (2001)

\bibitem{Wal01a}
Waller, B.N.: Classifying and analyzing analogies. Informal Logic  21(3),
  199--218 (2001)

\bibitem{Whi10a}
Whittaker, J.H.: The logic of authoritative revelations. International Journal
  for Philosophy of Religion  68(1--3),  167--181 (2010)

\end{thebibliography}


\end{document}